\documentclass[11pt,a4paper]{amsart}
\usepackage[margin=25mm]{geometry}
\usepackage[numbers]{natbib}
\usepackage{hyperref}

\usepackage{amssymb,amsmath}
\usepackage{amsthm}
\usepackage{bbm, bm}
\usepackage{stmaryrd}
\usepackage{color}
\usepackage{graphicx}
\usepackage{caption}
\usepackage{float}
\usepackage{wrapfig}
\usepackage{subcaption}
\usepackage{amsaddr}
\usepackage[mathscr]{eucal}
\usepackage{enumerate}
\usepackage{orcidlink}
\usepackage{lineno}
\usepackage{abstract}

\input xy
\xyoption{all} %\tolerance=500

\numberwithin{equation}{section}

\theoremstyle{definition}

\newenvironment{remark}[1][Remark.]{\begin{trivlist}
\item[\hskip \labelsep {\bfseries #1}]  }{ \end{trivlist}}

\newcommand{\Diffeo}{\mathrel{\raisebox{-.25ex}{$\xrightarrow{\sim}$}}}

\newcommand{\Id}{\mathbbmss{1}}

\newcommand{\rmd}{\textnormal{d}}

\DeclareMathOperator{\End}{End}
\DeclareMathOperator{\Vect}{Vect}

\DeclareMathOperator{\Span}{Span}

\DeclareMathOperator{\Der}{Der}

\font\black=cmbx10 \font\sblack=cmbx7 \font\ssblack=cmbx5 \font\blackital=cmmib10  \skewchar\blackital='177
\font\sblackital=cmmib7 \skewchar\sblackital='177 \font\ssblackital=cmmib5 \skewchar\ssblackital='177
\font\sanss=cmss10 \font\ssanss=cmss8 %scaled 900
\font\sssanss=cmss8 scaled 600 \font\blackboard=msbm10 \font\sblackboard=msbm7 \font\ssblackboard=msbm5
\font\caligr=eusm10 \font\scaligr=eusm7 \font\sscaligr=eusm5  \font\fraktur=eufm10
\font\sfraktur=eufm7 \font\ssfraktur=eufm5 
\font\bsymb=cmsy10 scaled\magstep2
\def\all#1{\setbox0=\hbox{\lower1.5pt\hbox{\bsymb
       \char"38}}\setbox1=\hbox{$_{#1}$} \box0\lower2pt\box1\;}
\def\exi#1{\setbox0=\hbox{\lower1.5pt\hbox{\bsymb \char"39}}
       \setbox1=\hbox{$_{#1}$} \box0\lower2pt\box1\;}

\def\tx#1{{\fam0\relax#1}}

\newfam\bifam
\textfont\bifam=\blackital \scriptfont\bifam=\sblackital \scriptscriptfont\bifam=\ssblackital

\newfam\blfam
\textfont\blfam=\black \scriptfont\blfam=\sblack \scriptscriptfont\blfam=\ssblack

\newfam\bbfam
\textfont\bbfam=\blackboard \scriptfont\bbfam=\sblackboard \scriptscriptfont\bbfam=\ssblackboard

\newfam\ssfam
\textfont\ssfam=\sanss \scriptfont\ssfam=\ssanss \scriptscriptfont\ssfam=\sssanss
\def\sss#1{{\fam\ssfam\relax#1}}

\newfam\clfam
\textfont\clfam=\caligr \scriptfont\clfam=\scaligr \scriptscriptfont\clfam=\sscaligr

\newfam\frfam
\textfont\frfam=\fraktur \scriptfont\frfam=\sfraktur \scriptscriptfont\frfam=\ssfraktur

\def\hpb#1{\setbox0=\hbox{${#1}$}
    \copy0 \kern-\wd0 \kern.2pt \box0}
\def\vpb#1{\setbox0=\hbox{${#1}$}
    \copy0 \kern-\wd0 \raise.08pt \box0}

\def\pmb#1{\setbox0\hbox{${#1}$} \copy0 \kern-\wd0 \kern.2pt \box0}
\def\pmbb#1{\setbox0\hbox{${#1}$} \copy0 \kern-\wd0
      \kern.2pt \copy0 \kern-\wd0 \kern.2pt \box0}
\def\pmbbb#1{\setbox0\hbox{${#1}$} \copy0 \kern-\wd0
      \kern.2pt \copy0 \kern-\wd0 \kern.2pt
    \copy0 \kern-\wd0 \kern.2pt \box0}
\def\pmxb#1{\setbox0\hbox{${#1}$} \copy0 \kern-\wd0
      \kern.2pt \copy0 \kern-\wd0 \kern.2pt
      \copy0 \kern-\wd0 \kern.2pt \copy0 \kern-\wd0 \kern.2pt \box0}
\def\pmxbb#1{\setbox0\hbox{${#1}$} \copy0 \kern-\wd0 \kern.2pt
      \copy0 \kern-\wd0 \kern.2pt
      \copy0 \kern-\wd0 \kern.2pt \copy0 \kern-\wd0 \kern.2pt
      \copy0 \kern-\wd0 \kern.2pt \box0}

\mathchardef\za="710B  %\alpha
\mathchardef\zb="710C  %\beta
\mathchardef\zg="710D  %\gamma
\mathchardef\zd="710E  %\delta
\mathchardef\zve="710F %\epsilon
\mathchardef\zz="7110  %\zeta
\mathchardef\zh="7111  %\eta
\mathchardef\zvy="7112 %\theta
\mathchardef\zi="7113  %\iota
\mathchardef\zk="7114  %\kappa
\mathchardef\zl="7115  %\lambda
\mathchardef\zm="7116  %\mu
\mathchardef\zn="7117  %\nu
\mathchardef\zx="7118  %\xi
\mathchardef\zp="7119  %\pi
\mathchardef\zr="711A  %\rho
\mathchardef\zs="711B  %\sigma
\mathchardef\zt="711C  %\tau
\mathchardef\zu="711D  %\upsilon
\mathchardef\zvf="711E %\phi
\mathchardef\zq="711F  %\chi
\mathchardef\zc="7120  %\psi
\mathchardef\zw="7121  %\omega
\mathchardef\ze="7122  %\varepsilon
\mathchardef\zy="7123  %\vartheta
\mathchardef\zf="7124  %\varomega
\mathchardef\zvr="7125 %\varrho
\mathchardef\zvs="7126 %\varsigma
\mathchardef\zf="7127  %\varphi
\mathchardef\zG="7000  %\Gamma
\mathchardef\zD="7001  %\Delta
\mathchardef\zY="7002  %\Theta
\mathchardef\zL="7003  %\Lambda
\mathchardef\zX="7004  %\Xi
\mathchardef\zP="7005  %\Pi
\mathchardef\zS="7006  %\Sigma
\mathchardef\zU="7007  %\Upsilon
\mathchardef\zF="7008  %\Phi
\mathchardef\zW="700A  %\Omega
\mathchardef\zC="7009  %\Psi

\newcommand{\be}{\begin{equation}}
\newcommand{\ee}{\end{equation}}

\newcommand{\bea}{\begin{eqnarray}}
\newcommand{\eea}{\end{eqnarray}}
\def\*{{\textstyle *}}
\newcommand{\R}{{\mathbb R}}

\newcommand{\C}{{\mathbb C}}
\newcommand{\Z}{{\mathbb Z}}

\newcommand{\s}{{\textstyle *}}

\def\Vect{\sss{Vect}}

\def\xi{\tx{i}}

\def\s*{{\scriptstyle *}}

\newcommand{\beas}{\begin{eqnarray*}}
\newcommand{\eeas}{\end{eqnarray*}}
\title{The Carrollian Superplane and Supersymmetry} 
\author{Andrew James Bruce \,\orcidlink{0000-0001-8197-2263} } 
  \email{andrewjamesbruce@googlemail.com}

   \date{\today}
 
\begin{document}
 %%%%%%%%%%%%%%%%%%%%%%%%%%%
 \maketitle
%%%%%%%%%%%%%%%%%%%%%%%%%%%
\vspace{-20pt}
%%%%%%%%%%%%%%%%%%%%
\begin{abstract}{\noindent{This note provides an intrinsic construction of the Carrollian superplane $\Pi \mathbb{S}\simeq \R^{2|4}$ as a supermanifold generalisation of the Carrollian plane. Moving away from the $c\rightarrow 0$ limit of relativistic spinors, we define Carroll spinors as sections of a degenerate Clifford module. We show that the Carrollian superplane is a principal $\R^{1|2}$-bundle. Once clock forms and a complementary basic one-form are specified, there is a pair of odd vector fields that generate novel $N =2$ Carrollian supersymmetry transformations, not all of which come from an Inönü--Wigner contraction of a Poincaré superalgebra.  }  }\\
\noindent {\Small \textbf{Keywords:}{~Supermanifolds;~Carrollian Geometry;~Clifford Algebras:~Supersymmetry} \\
\noindent {\Small \textbf{MSC 2020:} \emph{Primary:}~58A50
~~\emph{Secondary:}~15A66;~53Z05}}
\end{abstract}
\medskip
\begin{flushright}
\emph{I dare say you never even spoke to Time! }\\
Lewis Carroll, Alice's Adventures in Wonderland,  (1865) 
\end{flushright}
\section{Introduction}
The notion of a Carrollian manifold, largely due to Duval et al. \cite{Duval:2014a,Duval:2014b,Duval:2014c},  is a manifold equipped with a degenerate metric whose kernel is spanned by a nowhere vanishing complete vector field. The earlier foundational works of Lévy-Leblond \cite{Lévy-Leblond:1965},  Sen Gupta \cite{SenGupta:1966}, and  Henneaux \cite{Henneaux:1979} must also be highlighted.  Null hypersurfaces, such as punctured future or past light-cones in Minkowski spacetime, and the event horizon of a Schwarzschild black hole, are examples of Carrollian manifolds. A large part of the renewed interest in such manifolds is their role in flat space holography:  the boundary theory lives on future null infinity $\mathscr{I}^+ \cong \mathbb{R} \times S^{d-2}$, which comes with a Carrollian structure, for example \cite{Bagchi:2024,Ruzziconi:2026} and references therein. Carrollian physics has attracted a lot of attention from various perspectives, including hydrodynamics and condensed matter physics--for a review, the reader may consult Bagchi et al. \cite{Bagchi:2025}.  \par
Carrollian manifolds are often considered as the non-relativistic limit $c \rightarrow 0$ of Lorentzian manifolds; loosely, the Carrollian limit removes the temporal components of a metric, leaving only a degenerate metric with a rank-1 kernel. However, viewing Carrollian manifolds as just `broken' Lorentzian manifolds can obscure the intrinsic geometry.  For example, Carrollian manifolds naturally have a principal $\R$-bundle structure, as highlighted by Ciambelli et al. \cite{Ciambelli:2019}. \par 
In this note, we construct the Carroll superplane $\Pi \mathbb{S} \simeq \R^{2|4}$, as a supermanifold generalisation of the Carroll plane.  In particular, the Grassmann odd coordinates transform as Carroll spinors, which we carefully define using the degenerate Clifford algebra of the Carroll plane. We will refer to this degenerate Clifford algebra as a Carroll--Clifford algebra. The approach taken is to construct the geometry intrinsically rather than as a limiting procedure of a standard superspace.  \par 
We show that the Carroll superplane is a principal $\R^{1|2}$-bundle (where the group product is the abelian product given by shifts) equipped with a degenerate invariant metric whose kernel is spanned by the vertical vector fields.  Via selecting an Ehresmann connection, we construct the clock forms, which are a triple of invariant one-forms associated with the one even and two odd times.  Moreover, once a basic one-form has been chosen, a geometric (equivariant) form of supersymmetry can be constructed; this is reminiscent of the geometric formulation of supergravity. In particular, the `extra data' of a connection and a one-form are required and considered to be part of the intrinsic geometry. Upon the assumption that the clock forms are closed, we obtain the expected form of Carrollian supersymmetry (also referred to as C-supersymmetry \cite{Koutrolikos:2023}), schematically   
$$\{Q,Q\} = \Psi(x) \, \partial_t\,.$$
Note, however, the `structure constants' are now, in general, functions of position--thus we have a Lie--Rinehart pair rather than a Lie algebra.  There is a lot of freedom in the basic one-form, and it may be chosen constant, which is consistent with an Inönü--Wigner contraction of a Poincaré superalgebra.  The $x$-dependence of the basic one-form/structure functions is at odds with Poincaré symmetry, and thus not all Carrollian supersymmetries have a relativistic parent. In other words, the class of Carrollian supersymmetries is larger than just those that arise from the $c\rightarrow 0$ limit.\par   
We remark that a similar, but distinct approach to Carrollian spinors was given by  Bagchi et al. \cite{Bagchi:2023}, Banerjee et al. \cite{Banerjee:2023} \& Grumiller et al. \cite{Grumiller:2026}. In particular, the only metric-like structure on the Carroll plane is the degenerate metric, and so the degenerate Clifford algebra module is generated by gamma matrices with lower indices only.  The limit approach to fermions was carefully discussed by Bergshoeff \cite{Bergshoeff:2024}. Carrollian conformal superalgebras were studied by Zheng \& Chen \cite{Zheng:2025}, and they provided novel algebras that are not constructed as the $c\rightarrow 0$ limit of a Poincaré superalgebra. Thus, the results of this note sit comfortably with Zheng \& Chen's constructions, although they work in $d= 3,4$.  Concha \& Ravera \cite{Concha:2025} classify kinematical superalgebras via semigroup expansions (S-expansions) in their study of non-Lorentzian supergravity theories. Supersymmetric Carrollian theories, using $c\rightarrow 0$ limits, were studied by Kasikci et al. \cite{Kasikci:2024} and Koutrolikos \& Najafizadeh \cite{Koutrolikos:2023}.  The foundations of supersymmetric Carrollian holography were established through the study of super-BMS algebras by Bagchi et al. \cite{Bagchi:2022} and the derivation of superconformal Carrollian holograms from AdS/CFT by Lipstein et al. \cite{Lipstein:2025}. \par 
While the field of Carrollian supersymmetry is rapidly evolving, it is fair to say the subject is relatively poorly understood.  Much like the Lorentzian case, supersymmetric Carrollian theories are expected to have better renormalisation properties. For instance, the boson-fermion cancellation mechanism may lead to well-defined asymptotic charges which otherwise are divergent for purely bosonic theories.  Within the flat space holography programme, Carrollian supersymmetry may place constraints on the dual theories, restricting the admissible classes of theories on the boundary.  \par 
\medskip 
\noindent\textbf{Our use of supermanifolds:} We understand supermanifolds as locally superringed spaces, see for example Carmeli \cite{Carmeli:2011}, and we will assume a working knowledge of supergeometry. However, we will mostly work using coordinates and not employ deep results from the general theory of supermanifolds. Principal bundles in the category of supermanifolds are covered by Bartocci et al. \cite[Chapter VII]{Bartocci:1991}. We will denote the Garssmann parity of an object by `tilde', i.e.,  $\widetilde{O} \in \Z_2 = \{0,1 \}$.
%
%%%%%%%%%%%%%%%%%%%%%%%%%%%%%%%%%%%%
%
\section{The Carroll Plane and Carroll Superplane}
\subsection{The Canonical Carrollian Geometry of the Plane}
The underlying smooth manifold structure is $M \cong \R^2$, which we equip with standard global coordinates $(t,x)$. The canonical (weak) Carrollian structure on $M$ is given by $g = \delta x \otimes \delta x$ and $\kappa = \partial_t$. The reader can directly observe that $\ker(g) = \Span \{\kappa \}$.  The admissible changes of coordinates we consider are the \emph{extended Carroll transformations} 
\begin{align}\label{eqn:ConCarTran}
& t' = t -  \alpha\, f(x)\,, && x' = x - \beta \,,
\end{align}
where $\alpha$ and $\beta$ are constants with units of time and length, respectively. Here $f \in C^\infty(\R)$ is an arbitrary smooth function. The specific transformations $t' = t - \alpha\, f(x)$ are referred to as \emph{supertranslations}, and show that the isometry group here is infinite dimensional; the symmetries we consider are BMS-like. Note that Taylor expanding the smooth function, we obtain temporal shifts and Carrollian boost as the degree zero and one terms.  That is, the group defined by \eqref{eqn:ConCarTran} includes the standard Carrollian group $\mathsf{Carr}_{2}$, see Lévy-Leblond \cite{Lévy-Leblond:1965} \par 
A direct calculation gives 
\begin{equation}\label{eqn:ParDerChange}
\partial_ {t'} = \partial_t\,, \qquad  \partial_{x'}  = \partial_x+ \big(  \alpha \,\partial_x f(x)\big)\partial_t\,,
\end{equation}
meaning that $\kappa = \partial_t$ is invariant under the extended Carroll transformations.  In short, the Carrollian structure $(g, \kappa)$ is preserved under \eqref{eqn:ConCarTran}. \par 
Observe that we have a principal $\R$-bundle $\mathrm{prj}: M \rightarrow \Sigma\cong \R$, where the right action in coordinates is  $(t, x)\triangleleft r = (t+r, x)$. The vertical transformation part of the extended Carroll transformations \eqref{eqn:ConCarTran} we thus interpret as gauge transformations. Moreover, the degenerate metric is right-invariant, or in more physical language, the metric is stationary. 
%
%%%%%%%%%%%%%%%%%%%%%%%%%%%%%%%%%%%%
%
\subsection{The Carroll--Clifford Algebra}
For an overview of the theory of Clifford algebras, the reader may consult Lounesto \cite{Lounesto:2001}. We will fix a field $\mathbb{F} := \R$ or $\C$.  We define the Carroll--Clifford algebra $\mathcal{CC}l(\mathbb{F}):=\mathcal{C}l_{1,0,1}(\mathbb{F})$ as the associative $\mathbb{F}$-algebra with generators $\big\{\Id, e_t, e_x   \big\}$, subject to the relations
\begin{equation}
\{e_t, e_t \} =0\, , \qquad \{e_t, e_x \}=0\, , \qquad \{e_x, e_x \}= 2 \, \Id\,.
\end{equation}
Thus, a general element of the Carroll--Clifford algebra is of the form
$$A = a\, \Id + b\, e_t + c\, e_x + d\, e_te_x\,,$$ 
with $a,b,c$ and $d \in \mathbb{F}$. Note that we have what is referred to in the literature as a degenerate Clifford algebra. \par
A key observation here is that we have a canonical algebra isomorphism
$$\phi : \mathcal{CC}l(\mathbb{F}) \Diffeo \Lambda^1(\mathbb{F})\otimes_{\mathbb{F}}\mathcal{C}l_{1,0,0}(\mathbb{F})\,,$$ 
given by the associative algebra map 
\begin{equation}
\phi(\Id) := \Id \otimes_{\mathbb{F}}  \Id\, , \qquad \phi(e_t) := \theta \otimes_{\mathbb{F}}\Id\,, \qquad \phi(e_x) := \Id \otimes_{\mathbb{F}}e_x\,. 
\end{equation}
Neglecting the explicit reference to the isomorphism and tensor product, we will write
$$\mathcal{CC}l(\mathbb{F}) \ni A = a + b\, \theta + c \, e_x + d \, \theta  e_x\,.$$ 
The Carroll--Clifford algebra comes with a natural $\Z_2 \times \Z_2$-grading defined on the generators as 
\begin{equation}
\deg(\Id) = (0,0)\,, \qquad  \deg(\theta) = (1,0)\, , \quad \deg(e_x) = (0,1)\,.
\end{equation}
\begin{remark} The assignment of the grading is unique; the degrees of $\theta$ and $e_x$  may be exchanged.  It is also important to note $\mathcal{CC}l(\mathbb{F})$ is not a $\Z_2 \times \Z_2$-graded commutative algebra. 
\end{remark}
The \emph{Carrollian boost generator} is defined as 
\begin{equation}
S_{tx} := \frac{1}{4} \, [\theta, e_x] = \frac{1}{2} \theta e_x \, (= - S_{xt})\,.
\end{equation}
Note that due to the nilpotency of $\theta$,   $(S_{tx})^2 =0$. Directly, we observe that 
\begin{align*}
&[S_{tx}, \theta] = \frac{1}{2}\big( \theta e_x \theta - \theta^2 e_x  \big) = - \theta^2 e_x =0\,, \\
&[S_{tx}, e_x ] = \frac{1}{2} \big( \theta e_x^2 - e_x \theta e_x\big) = \theta e_x^2 = \theta\,.
\end{align*}
Comparing the above with \eqref{eqn:ParDerChange}, we see that we have a covariant vector representation of supertranslations, i.e., this ``mimics'' how partial derivatives transform.  
%
%%%%%%%%%%%%%%%%%%%%%%%%%%%%%%%%%%%%
%
\subsection{Carroll--Clifford Modules}
We will consider the vector space $\mathbb{F}^4$ as  $\Z_2$-graded and, more refined, $\Z_2 \times \Z_2$-graded;
$$\mathbb{F}^4 \ni \boldsymbol{v}= \begin{pmatrix}
    \boldsymbol{v}_0 \\
   \boldsymbol{v}_1
\end{pmatrix} =  \begin{pmatrix}
    v_{00} \\
    v_{11}\\
    v_{01}\\
    v_{10}
\end{pmatrix}\,.$$   
A \emph{Carroll--Clifford module} is defined as a $\Z_2 \times \Z_2$-grading preserving action
\begin{equation}
\rho : \mathcal{CC}l(\mathbb{F}) \longrightarrow \underline{\End}(\mathbb{F}^4)
\end{equation}
Respecting the $\Z_2 \times\Z_2$-grading places strict constraints on the form of the action.  In particular, employing the standard Pauli spin matrices
\begin{align}
\rho(\Id) = \begin{pmatrix}
    \Id & 0 \\
    0 & \Id
\end{pmatrix}\, , && \rho(\theta) = \begin{pmatrix}
    0 & \pm \sigma_+ \\
    \pm \sigma_+ & 0
\end{pmatrix}\, , && \rho(e_x) = \begin{pmatrix}
    0 & \sigma_3 \\
    \sigma_3 & 0\,.
\end{pmatrix}
\end{align}
A quick calculation gives
\begin{equation}
\rho(S_{tx})= \frac{1}{2}\begin{pmatrix}
    \mp \sigma_+ & 0 \\
    0 & \mp \sigma_+
\end{pmatrix}\,.
\end{equation}
We then observe that 
$$\rho(S_{tx}) \boldsymbol{v} = \begin{pmatrix}
    \mp v_{11} \\
    0\\
    \mp v_{10}\\
    0
\end{pmatrix}\,.$$
To build the Carroll spinor bundle, note that all the choices of $\rho(\theta)$ are similar; this essentially follows from the matrices all being rank $2$ and nilpotent. Different choices here will only result in different choices of fibre coordinates on the Carroll spinor bundle that differ by a sign.  \par 
The \emph{Carroll spinor bundle} on the Carrollian plane, we define as the vector bundle
$$\pi : \mathbb{S} \rightarrow M \cong \R^2\,,$$
with typical fibre $\mathbb{S}_p \Diffeo \R^4$ ($p \in \R^2$) and adapted coordinates $(t,x ; v_{00}, v_{11}, v_{01}, v_{10})$ together with admissible coordinate transformation  of the form 
\begin{subequations}
\begin{align}
&t' = t - \alpha\, f(x)\, && x' = x - \beta\,, \\ \label{eqn:sheara}
&v'_{00} = v_{00} + \frac{1}{2}\alpha \partial_x f(x)\, v_{11}\,, && v'_{11} = v_{11}\,,\\ 
&v'_{01} = v_{01} + \frac{1}{2}\alpha \partial_x f(x)\, v_{10}\,, && v'_{10} = v_{10}\,.\label{eqn:shearb}
\end{align}
\end{subequations}
The transformation rule for the fibre coordinates may be written as 
$$\boldsymbol{v}' = \boldsymbol{v} + \alpha \, \partial_x f(x)\, \rho(S_{tx})\boldsymbol{v} \,,$$
where the exact signs in the transformation rules are defined by the choice of $\rho(S_{tx})$. Note that as the matrix $\rho(S_{tx})$ is nilpotent, we do not need to exponentiate the infinitesimal transformation to obtain global ones, as is required in the Lorentzian case. Thus, the transformation rules \eqref{eqn:sheara} and \eqref{eqn:shearb} are shear transformations rather than rotations. While this realisation is already present in the $c\rightarrow 0$ approach to Carrollian spinors, this fact is not stressed enough.  Moreover, looking at the dimensions, $[\alpha] = [T]$, we see that 
\begin{equation}
\frac{[v_{00}]}{[v_{11}]} = \frac{[v_{01}]}{[v_{10}]} = \frac{[T]}{[L]}\,,
\end{equation}
that is, the ratios of the units must be `slowness', i.e, inverse speed.  As we do not have a fundamental speed, we cannot construct a meaningful dimensionless ratio.\par
It is important to note that the admissible coordinate transformations do not preserve the $\Z_2 \times \Z_2$-grading, but rather only the $\Z_2$-grading defined by total degree.  Thus, the Carroll spinor bundle is a $\Z_2$-graded vector bundle. As a graded vector bundle, we have the decomposition
$$\mathbb{S} \simeq \mathbb{S}_0 \oplus \mathbb{S}_1\,.$$   
Similarly, we define the complexified  Carroll spinor bundle $\mathbb{S}^{\C}$ by defining the typical fibres $(\mathbb{S}^\C)_p \Diffeo \R^4 \otimes_\R \C$.
%
%%%%%%%%%%%%%%%%%%%%%%%%%%%%%%%%%%%%
%
\subsection{The Carrollian superplane}
Using the Batchelor--Gawedzki theorem, we define the \emph{Carrollian superplane} as $\Pi \mathbb{S}$. Thus we have a supermanifold  equipped with coordinates  $(t,x,  \zeta^i, \eta^j)$, with $i,j \in \{1,2\}$, together with the admissible coordinate transformations 
\begin{align}\label{eqn:SuperTrans}
&t' = t - \alpha\, f(x) \, ,&& x' = x - \beta\,, \\ \nonumber
&\zeta^{i'} =\zeta^i + \frac{1}{2}\alpha \partial_x f(x)\, \eta^i\,, && \eta^{j'} = \eta^j\,.
\end{align}
We assign units to the Grassmann odd coordinates $[\zeta^i] = [T]$ and $[\eta^j] = [L]$.    
Observe that we have a principal $\R^{1|2}$-bundle structure, where the group structure on $\R^{1|2}$, given using the functor of points, is $(r_1 , \epsilon_1^i)\cdot (r_2 , \epsilon_2^i) := (r_1 + r_2 , \epsilon_1^i + \epsilon_2^i)$. That is, we consider the strictly abelian group structure of translations. The principal action can  be written, using the functor of points, as (ignoring the base coordinates)
\begin{equation}
(t, \zeta^i) \triangleleft (r, \epsilon^i) := ( t +  r, ~ \zeta^i + \epsilon^i ) \,,
\end{equation}
Where we assign units $[r] =[\epsilon^i]= [T]$. The reader can quickly verify that the coordinate changes are compatible with the action.  Moreover, as the coordinate changes are essentially shifts, the cocycle conditions are clearly satisfied.\par 
We then observe that the (canonical) right invariant degenerate metric on $\R^{2|4}$ is of the form 
\begin{equation}
g := \delta x \otimes \delta x \pm 2 \, \delta \eta^1 \otimes \delta \eta^2\,,
\end{equation}
where we have employed  the $\Z_2$-graded tensor product.  We then observe that the kernel is of rank $1|2$ and explicitly given by  
$$\ker(g) = \Span\big( \partial_t, \partial_{\zeta^1}, \partial_{\zeta^2} \big)\,.$$
We then observe that the structure underlying the Carroll superplane is the Carrollian plane. 
%
%%%%%%%%%%%%%%%%%%%%%%%%%%%%%%%%%%%%
%
\subsection{Connections and Clock Forms}
Directly, we observe that under the changes of coordinates \eqref{eqn:SuperTrans}, the partial derivatives transform as
\begin{subequations}
\begin{align}
& \partial_{t'} = \partial_t\,, && \partial_{x'} = \partial_x + \alpha\partial_xf(x) \partial_t - \frac{1}{2} \alpha \partial^2_x f(x) \eta^i \partial_{\zeta^i}\,,\\
& \partial_{\zeta^{i'}} = \partial_{\zeta^i}\,, &&  \partial_{\eta^{j'}} = \partial_{\eta^j} - \frac{1}{2} \alpha \partial_x f(x)\partial_{\zeta^j}\,.
\end{align}
\end{subequations}
Thus, as expected, we require an Ehresmann connection, which we refer to as the \emph{Carrollian connection} or \emph{Carrollian covariant derivative}
that realised the decomposition 
$$\Vect(\Pi \mathbb{S})\Diffeo \Vect_\mathsf{V}(\Pi \mathbb{S} ) \oplus \Vect_\mathsf{H}(\Pi \mathbb{S} )\, ,$$
into vertical and horizontal vector fields.  We thus define 
\begin{equation}
\nabla_x = \partial_x + \mathbb{A}_x ^{~t}\partial_t + \mathbb{A}_x^{~i}\partial_{\zeta^i}\,, \qquad \nabla_{\eta^j} = \partial_{\eta^j} + \mathbb{A}_j^{~k}\partial_{\zeta^k}\,,
\end{equation}
where the connection coefficients $\mathbb{A} = (\mathbb{A}_x^{~t}, \mathbb{A}_x^{~i} , \mathbb{A}_j^{~k})$, in general  depend on all the coordinates.   Under changes of coordinates, one can quickly deduce 
\begin{equation}
\mathbb{A}_x^{'~t}= \mathbb{A}_x^{~t} - \alpha \partial_x f\,, \qquad \mathbb{A}_x^{'~i} = \mathbb{A}_x^{~i} + \frac{1}{2} \alpha \partial_x^2 f \eta^i\,, \qquad \mathbb{A}_j^{'~k}=\mathbb{A}_j^{~k} + \frac{1}{2}\alpha \partial_x f \delta_j^{~k}\,.
\end{equation}
The invariant basis of one-forms, once a connection has been selected, gives the notion of the \emph{clock forms}
\begin{equation}
\tau^0 := \rmd t - \rmd x \mathbb{A}_x^{~t}\, , \qquad \tau^i := \rmd \zeta^i - \rmd x \mathbb{A}_x^{~i} - \rmd \eta^j \mathbb{A}_j^{~i}\,,  
\end{equation}
As a $\R^{1|2}$-valued one-form, we define the \emph{clock form} as  $\boldsymbol{\tau} := \mathrm{diag}(\tau^0, \tau^1, \tau^2)$.\par 
The \emph{Frobenius--Carroll curvature} of the connection is defined as 
\begin{align}\label{eqn:FrobCarrCurv}
R &: ~ \Vect_\mathsf{H}(\Pi \mathbb{S}) \times \Vect_\mathsf{H}(\Pi \mathbb{S}) \rightarrow \Vect_\mathsf{V}(\Pi \mathbb{S})\\ \nonumber 
& (X,Y) \longmapsto \mathrm{prj}_{\mathsf{V}} \big( [X,Y]\big)\,,
\end{align}
where $ \mathrm{prj}_{\mathsf{V}} : \Vect(\Pi \mathbb{S}) \rightarrow \Vect_\mathsf{V}(\Pi \mathbb{S})$ is the canonical projection. The curvature measures the non-closure of the horizontal distribution under the Lie bracket. A Carrollian connection is said to be \emph{flat}  if the Frobenius--Carroll curvature \eqref{eqn:FrobCarrCurv} vanishes, i.e., $R(X,Y) =0$ for all horizontal vector fields.\par 
As standard for connections, observe that $\tau^0(X) = 0$ and $\tau^i(X)=0$, if and only if $X \in \Vect_{\mathsf{H}}(\Pi \mathbb{S})$; the reader can quickly check this using local coordinates.  As, 
$$\rmd \omega (X,Y) =  X\big(\omega(Y) \big) -  (-1)^{\widetilde{X} \, \widetilde{Y}} \, Y \big( \omega(X) \big) - \omega ([X,Y])\,,$$
for all $\omega \in \Omega^1(\Pi \mathbb{S})$ and $X,Y \in \Vect_{\mathsf{H}}(\Pi \mathbb{S})$, we observe that
\begin{equation}
\rmd \tau^0(X,Y) = - \tau^0([X,Y])\,, \qquad \rmd \tau^i(X,Y) = - \tau^i([X,Y])\,.
\end{equation}   
Thus, $\rmd \boldsymbol{\tau} =0$ is equivalent to the Frobenius--Carroll curvature vanishing.\par
The clock form is said to be stationary if the Carrollian connection is a principal connection, i.e., we require $\Vect_{\mathsf{H}}(\Pi \mathbb{S})$ is equivariant. In local coordinates, this means that the components $\mathbb{A}$ are independent of $(t, \zeta^i)$. More invariantly, defining $\boldsymbol{\kappa}:= (\partial_t, \partial_{\zeta^1}, \partial_{\zeta^2})^T$, the stationary condition is $L_{\boldsymbol{\kappa}} \boldsymbol{\tau} =0$.\par 
Given a clock form, there is an underlying clock form on the Carrollian plane. To see this, observe that in any admissible coordinate system, the component of the connection $\mathbb{A}_x^{~i}$ is a collection of even (local) functions, thus it can be expanded as
$$\mathbb{A}_x^{~i}(t,x, \zeta, \eta) = \mathbb{A}_{0,x}^{~i}(t,x) + \textnormal{Terms Involving Odd Coordinates}\, ,$$  
thus $\mathbb{A}_{0,x}^{~i}(t,x)$ is a collection of local functions on $\R^2$ that transforms in the appropriate way. That is, we have a connection on the Carrollian plane. Moreover, $\mathbb{A}_x^{~i}$ is a collection of odd local functions, and so vanishes on the Carrollian plane.  If the Carrollian connection on $\Pi \mathbb{S}$, then the reduced Carrollian connection on $\R^2$ is also principal.
%
%%%%%%%%%%%%%%%%%%%%%%%%%%%%%%%%%%%%
%
\subsection{Geometric Supersymmetry}
Let us fix a principal Carrollian connection $\mathbb{A}$ and a basic (with respect to the principal bundle structure) 
$$\Psi = \rmd x \, \Psi_x(x, \eta) + \rmd \eta^i \, \Psi_i(x, \eta)\,.$$ 
By convention, $\Psi$ is taken as Grassmann odd, meaning that $ \Psi_x(x, \eta)$ is even and $ \Psi_i(x, \eta)$ is Grassmann odd.  Note that as a basic one-form, $\Psi$ is equivariant, and the components transform as scalars. Moreover, we will take the basic form to have units $[T]$, so the components have units $[T][L]^{-1}$.\par 
We then define the odd vector fields, which we refer to as the \emph{supergenerators}
\begin{equation}
Q_i := \nabla_{\eta^i} + \Psi_i \partial_t\,,
\end{equation} 
which is globally defined and is invariant under coordinate transformations.  Directly we calculate
\begin{equation}
\{Q_i, Q_j \} =  \big( \partial_{\eta^i}\Psi_j + \partial_{\eta^j}\Psi_i \big)\partial_t+\big( \partial_{\eta^i}\mathbb{A}_j^{~k} + \partial_{\eta^j}\mathbb{A}_i^{~k} \big)\partial_{\zeta^k} \in \Vect_{\mathsf{V}}(\Pi \mathbb{S})\,.
\end{equation}
Note that $Q^2 \sim \partial_t  + \partial_\zeta$, is not quite of generally expected form for a version of Carrollian supersymmetry. Specifically,  in more standard approaches, the Carrollian supercharge squares to temporal translations, see for example \cite{Koutrolikos:2023}. Observe that if the Carrollian connection is flat, then we obtain 
\begin{equation}\label{eqn:SusyBracket}
\{Q_i, Q_j \} =  \big( \partial_{\eta^i}\Psi_j + \partial_{\eta^j}\Psi_i \big)\partial_t= 2 \, \Psi_{(ij)}(x) \, \partial_t\,,
\end{equation}
where $\Psi_i(x, \eta) = \eta^k\Psi_{ki}(x)$ due to  $\Psi_i$ being Grassmann odd, and we define the symmetrisation as standard, i.e.,  $2 \, \Psi_{(ij)} = \Psi_{ij} + \Psi_{ji}$. Note that one may choose each $\Psi_{ij}$ and so each $\Psi_{(ij)}$ to be constants. In particular, one may choose $\Psi_{(ij)}= \delta_{ij}$, so that the \eqref{eqn:SusyBracket} is of the classical form.   Even with non-constant components $\Psi_i$, observe that  $[Q_i, \{Q_j, Q_k\}]=0$, which is important when defining the associated Lie--Rinehart structure.
\par
The associated supersymmetry transformations expressed via coordinates are thus 
\begin{align}\label{eqn:SusyTrans}
& t \mapsto t + \zx^i \Psi_i(x, \eta)\,, && x \mapsto x\,, \\ \nonumber 
&  \zeta^i \mapsto \zeta^i + \zx^k\mathbb{A}_k^{~i}(x, \eta)\, , && \eta^j \mapsto \eta^j +\zx^j\,,
\end{align}
where $\zx^i$  are Grassmann odd parameters with units $[\eta] = [L]$. Note that these supertransformations respect the principal bundle structure. Specifically, they are compatible with the $\R^{1|2}$-principal action.\par 
Restricting to Carrollian boosts for simplicity, i.e., $f = x$, we define the vector fields 
\begin{equation}
H = \partial_t\,, \quad  P = \nabla_x\,, \quad B = x \partial_t - \frac{1}{2}\eta^i \partial_{\zeta^i} \, , \quad  Q_j = \nabla_{\eta^j} + \Psi_j \partial_t\,, \quad X_k = \partial_{\zeta^k}\,,
\end{equation} 
and observe that the non-zero Lie brackets are 
\begin{align}\label{eqn:NonVanBrack}
&[P, B] = H\,, && [P, Q_i] = P(\Psi_i)\,H \,, \\   \nonumber 
&[B, Q_j]= \frac{1}{2} X_j\,, && \{Q_k , Q_l \} = \big(Q_k(\Psi_l)+ Q_l(\Psi_k) \big)\, H\,.
\end{align}
Recall that a  Lie--Rinehart superpair is a pair $(\mathcal{A}, V)$ where $\mathcal{A}$
is an associative supercommutative algebra, and $V$ is a Lie superalgebra such that
$V$ is an $\mathcal{A}$-module, and $V$ acts as derivations on $\mathcal{A}$. In particular, $[u, \psi \, v ] = \mathsf{a}(u)\psi \, v + (-1)^{\widetilde{\psi}\, \widetilde{u}} \, \psi \, [u,v]$, where $\mathsf{a} : V \rightarrow \Der(\mathcal{A})$ is referred to as the anchor map. For details, the reader may consult \cite{Huebschmann:2004} and references therein. \par 
The \emph{Carroll--Lie--Rinehart superpair} is defined as follows. The associative supercommutative algebra is $\mathcal{A} := C^\infty(\Pi\mathbb{S} )$, and  the Lie algebra is $V:=\Span\{ H, P, B, Q_i , X_j\}$, together with their Lie brackets \eqref{eqn:NonVanBrack}. The anchor map $\mathsf{a} : V \rightarrow \Vect(\Pi \mathbb{S})$ is given by $\mathsf{a}(Z)\psi:= Z(\psi)$, for all $Z \in V$ and $\psi \in C^\infty(\Pi\mathbb{S} )$.\par 
To define the supercovariant derivatives in the case of closed clocks, we require $\Psi_{ij}= \Psi_{(ij)}$ and we write $Q_i = \nabla_{\eta^i} + \eta^j \Psi_{(ji)}(x)\partial_t$.  The \emph{supercovariant derivatives} are defined as  
\begin{equation}
D_i := \nabla_{\eta^i} - \eta^j \Psi_{(ji)}(x)\partial_t\,.
\end{equation}
The reader can quickly verify that
$$\{ D_i, D_j\}= - 2 \, \Psi_{(ij)}(x)\partial_t\,, \qquad \{D_i, Q_j \} = (\Psi_{(ij)}- \Psi_{(ji)})\partial_t = 0\,.$$ 
%
%%%%%%%%%%%%%%%%%%%%%%%%%%%%%%%%%%%%
%
\section{Concluding Remarks}
In this note, we have, via the intrinsic construction of Carrollian spinors, built a supermanifold generalisation of the Carrollian plane $(\Pi \mathbb{S}, g, \boldsymbol{\kappa})$. Once (equivariant) clocks have been chosen $\boldsymbol{\tau}$, alongside a complementary basic one-form $\Psi$, then a geometric supersymmetry can be constructed. The full structure should then be considered  to be $(\Pi \mathbb{S}, g, \boldsymbol{\kappa}, \boldsymbol{\tau}, \Psi)$. The form of the supersymmetry, specifically  
$$\{Q_i, Q_j \} =  2 \, \Psi_{(ij)}(x) \, \partial_t\,,$$
is novel and shows that not all Carrollian supersymmetries can be constructed via a $c\rightarrow 0$ procedure.  In particular, the $x$-dependence demonstrates that these supersymmetries and not, in general, an Inönü--Wigner contraction of a Poincaré superalgebra.\par
It is possible to consider a ``smaller'' notion of the Carrollian superplane by considering $\Pi \mathbb{S}_0\simeq \Pi \mathbb{S}_1$. In this case, we have a principal $\R^{1|1}$-bundle, and the constructions in this note specialise to that setting directly. However, we do not have a degenerate metric whose kernel is spanned by the vertical vector fields.  As the classical notion of a Carrollian manifold requires a degenerate metric, we accept that we require four Grassmann odd coordinates. \par  
While the focus of this note has been mathematics, specifically degenerate Clifford algebras, principal bundles and supermanifolds, it is expected that the constructions presented here will be of some use in building intrinsic Carrollian supersymmetric field theories.   For instance, considering superstatic (scalar) fields, so superfields $\Phi$ that satisfy $X_i(\Phi)=0$. In coordinates, we have 
$$\Phi = \phi(t,x) + \eta^i \psi_i(t,x) + \frac{1}{2} \eta^i \eta^j F_{ji}(t,x)\,,$$
which have their standard interpretation as a pair of even fields and a pair of odd fields: note $F_{12} = - F_{21}$ are the only non-vanishing components of $F_{ji}$.  The component-wise Carrollian supertransformations are defined as $\delta_{\zx} \Phi := \zx^i Q_i \Phi$, and explicitly are  
$$\delta_\zx \phi = \zx^i\psi_i\,, \qquad \delta_\zx \psi_j = \zx^k\big(\Psi_{(kj)}  \partial_t \phi - F_{kj} \big )\,, \qquad \delta_\zx F_{ij} = - \zx^k\big(\Psi_{(ki)}\partial_t\psi_j  - \Psi_{(kj)}\partial_t\psi_i\big)\,.$$
More generally, one can mimic standard superspace methods to construct invariant actions. The coordinate Berezin volume is invariant under the coordinate changes \eqref{eqn:SuperTrans} and the Carrollian supersymmetry transformations \eqref{eqn:SusyTrans}. We can then define actions 
$$ \mathsf{S}[\Phi] = \int  \mathcal{L}(\Phi, D\Phi)\,, $$
which are automatically invariant under Carrollian supersymmetry. Written out in components, the Euler--Lagrange equations will be of `electric-type', i.e., contain only partial derivatives with respect to time. Thus, although quite exotic Lagrangian densities can be constructed, they cannot possess on-shell propagating degrees of freedom.  

%
%%%%%%%%%%%%%%%%%%%%%%%%%%%%%%%%%%%%
%
%\section*{Acknowledgements}  

%
%%%%%%%%%%%%%%%%%%%%%%%%%%%%%%%%%%%%
%

%%%%%%%%%%%%%%%%%%%%%%%%%%%%%%%%%%
\end{document}